\documentclass[journal]{IEEEtran}
\usepackage{amsmath,amssymb}
\newtheorem{theorem}{Theorem}
\newtheorem{example}{Example}
\begin{document}
\title{Subset Codes for Packet Networks}%
\author{Mladen~Kova\v cevi\' c,~\IEEEmembership{Student Member,~IEEE,}
        and Dejan~Vukobratovi\' c,~\IEEEmembership{Member,~IEEE}%
\thanks{Manuscript received October 28, 2012; revised January 15, 2013.}%
\thanks{This work was supported by the Ministry of Science and Technological
        Development of the Republic of Serbia (grants No.\ TR32040 and III44003).\newline
        D.\ Vukobratovi\'c was partly supported by the provincial government of 
        Vojvodina, Republic of Serbia (grant No.\ 114-451-2061/2011-01).}%
\thanks{The authors are with the Department of Power, Electronics, and
        Communication Engineering, University of Novi Sad, Serbia.
        E-mails: \{kmladen, dejanv\}@uns.ac.rs.}}%
\maketitle
\begin{abstract}
In this paper, we present a coding-theoretic framework for message transmission
over packet-switched networks. Network is modeled as a channel which can induce
packet errors, deletions, insertions, and out of order delivery of packets. The
proposed approach can be viewed as an extension of the one introduced by K\"otter
and Kschischang for networks based on random linear network coding. Namely, while
their framework is based on \emph{subspace} codes and designed for networks in
which network nodes perform random linear combining of the packets, ours is based
on the so-called \emph{subset} codes, and is designed for networks employing
\emph{routing} in network nodes.
\end{abstract}
\section{Introduction}
\IEEEPARstart{P}{acket-switched} networks that employ routing as a means
for transmitting packets between pairs of users are in widespread use in
communications today \cite{galag}. We formulate here a framework for end-to-end
forward error correction in such networks. We are motivated by the work of
K\"otter and Kschischang \cite{kk} in which the authors define so-called \emph{
subspace codes} and show that these codes, and particularly their constant-dimension
versions, are adequate constructions for error and erasure recovery in networks
employing random linear network coding (RLNC). The two frameworks turn out to
be similar in many respects. Indeed, most concepts defined in our model have
natural analogs in the subspace coding setting. On the other hand, there are
some important differences between the two models, one of which will lead to
a surprising conclusion that the codes for packet networks that are introduced
here are equivalent (in a certain sense) to the classical binary codes in the
Hamming space.
\par Let us now informally state the basic idea behind both approaches. Consider
a network, abstracted as a communication channel, that acts on the transmitted
packets by some randomized transformation (not including errors, erasures, etc.).
In the case of RLNC networks, the channel transformation represents random linear
combining of the source packets. In the case of networks based on routing, the
transformation corresponds to permuting the packets in an unpredictable, and
essentially random way. Namely, in such networks the packets with the same
destination are frequently sent over different routes in the network and, as a
consequence, they are received in practically arbitrary order (see the following
section for a more detailed discussion of the channel model). The idea of sending
information through such channels is very simple: \emph{Encode the information in
an object that is invariant under the given transformation}. This has led K\"otter
and Kschischang to the abstraction of the channel corresponding to RLNC networks
(the operator channel) and the definition of codes for such a channel. In this
case, the object invariant under random linear combinations of the packets
is the vector space spanned by those packets\footnote{\,Strictly speaking, it is
invariant only with high probability -- if the transformation is full-rank.}.
Hence, the ``codewords" are in this context taken to be subspaces of some ambient
vector space \cite{kk}.
\par In the case under consideration here, namely routed packet networks, we
need an object that is invariant under random permutations of the packets. Such
an object is a \emph{set}. Therefore, a natural idea is to consider \emph{sets
of packets} as ``codewords" in this context. If $S$ is the set of all possible
packets, the appropriate space in which such codes are to be defined is the set
of all subsets of $S$, denoted ${\mathcal P}(S)$. In the following, we provide
precise definitions and properties of the codes in ${\mathcal P}(S)$, which are
proposed as relevant to the problem of reliable data transmission over routed
packet networks.
\section{The system model}
\label{channel}
Consider a packet-switched network in which a source node wishes to communicate
with a destination node (or with multiple destination nodes). We assume that a
message to be sent consists of a batch of packets (also called a generation) that
are ``simultaneously" injected into the network. Due to varying topology and load,
the packets from the same batch can be sent over different routes in the network
and their order is in general not preserved at the receiving side. This is especially
true for, e.g., mobile ad-hoc networks where the topology is rapidly changing, and
heavily loaded datagram-based networks in which the packets are frequently redirected
in order to balance the load over different parts of the network. Hence, we will
model networks as \emph{packet permutation channels} which can deliver injected
packets in an arbitrary order at the destination. Apart from permutations, there are
various other unwanted effects the network can impose on the transmitted packets.
We consider here three of them: \emph{errors, deletions, and insertions}. Errors
are random alterations of packet symbols caused by noise, malfunctioning of network
equipment, etc. Packet deletions correspond to the fact that some packets can be
``lost" in the channel, in which case the receiver is unaware of them being sent\footnote{\,In
the networking literature, the term ``erasures" is also used in this context. We
will use the term ``deletions" since it is more appropriate from the coding theory
viewpoint.}. They can occur for many reasons, finite buffering capabilities of
routers, router/link failures, etc. Finally, packet insertions are a form of
malicious behavior, where some user imitates the true source of the data, and
wants the receiver to misinterpret the data.
\par In the following section we will introduce \emph{subset codes} as adequate
for the above-described network model, i.e., for the permutation channel with
errors, deletions, and insertions. Given that these codes are, as already noted,
defined on the power set of the set of all possible packets $S$, we can also
give a more formal definition of the considered channel: It is a discrete memoryless
channel with input and output alphabets equal to ${\mathcal P}(S)$. The channel
is completely described by its transition probabilities (the probabilities of
mapping the input subset $X$ to the output subset $Y$, for all $X,Y\in{\mathcal P}(S)$)
which, on the other hand, are determined by the joint statistics of errors, deletions,
and insertions of the elements of $S$.
\par As a final remark in this section, we emphasize that this paper considers
an end-to-end network transmission model. Hence, it is implicitly assumed that
(subset) coding is done on the transport or application layer.
\section{Codes for packet networks}
\label{codes}
\subsection{Power sets and subset codes}
\label{subset}
Let $S$ be a nonempty finite set, and let ${\mathcal P}(S)$ denote the power
set of $S$, i.e., the set of all subsets of $S$. A natural metric associated
with this space is:
\begin{equation}
\label{metricI}
  d(X,Y) = |X\bigtriangleup Y|
\end{equation}
for $X,Y\in{\mathcal P}(S)$, where $\bigtriangleup$ denotes the symmetric
difference of sets. It can also be written as $d(X,Y) = |X\cup Y| - |X\cap Y| =
|X| + |Y| - 2|X\cap Y| = 2|X\cup Y| - |X| - |Y|$. This distance is the length
of the shortest path between $X$ and $Y$ in the Hasse diagram \cite{lattice}
of the lattice of subsets of $S$ ordered by inclusion. It is analogous to the
subspace metric defined in \cite{kk}. This diagram plays a role similar to the
Hamming hypercube for the classical codes in the Hamming metric (actually, it
is isomorphic to the Hamming hypercube, see Section \ref{isometry}). Another
useful metric is given by:
\begin{equation}
\label{metricII}
  d'(X,Y) = \max\{|X\setminus Y| , |Y\setminus X|\}.
\end{equation}
It can also be written as
$d'(X,Y) = \max\{|X|,|Y|\} - |X\cap Y| = |X\cup Y| - \min\{|X|,|Y|\}$,
and it is analogous to the injection metric for subspace codes \cite{silva}.
Direct proofs that $d$ and $d'$ are indeed metrics are easy and very similar
to the proofs for subspace and injection metrics, and we shall therefore omit
them. In the following, we will only use distance $d$ and refer to it as the
\emph{subset metric}.
\par One can define codes in the space ${\mathcal P}(S)$ in the usual way.
Namely, a \emph{subset code} $\mathcal C$ is simply a nonempty subset of
${\mathcal P}(S)$. Important parameters of such a code are its cardinality,
$|{\mathcal C}|$, minimum distance:
\begin{equation}
 \min_{X,Y\in {\mathcal C},\;X\neq Y} d(X,Y),
\end{equation}
maximum cardinality of the codewords:
\begin{equation}
 \max_{X\in {\mathcal C}} |X|,
\end{equation}
and the cardinality of the ambient set, $|S|$. If ${\mathcal C}\subseteq{\mathcal P}(S)$
has minimum distance $d$, and every codeword is of cardinality at most $\ell$,
we say that it is a code of type $[\log |S|,\log |{\mathcal C}|,d;\ell]$ (the
base of the logarithm is generally arbitrary; we will assume that it is 2, and
hence that the lengths of the messages are measured in bits).
If all codewords of ${\mathcal C}$ are of cardinality $\ell$, we say that it
is a constant-cardinality code. A significant advantage of constant-cardinality
codes is that the receiver knows in advance how many packets it needs to receive
in order to initiate decoding, similarly to the constant-dimension codes in
projective spaces \cite{kk}. The rate of an $[n,k,d;\ell]$ code is defined by:
\begin{equation}
  R=\frac{k}{n\ell}.
\end{equation}
\par In the intended application of subset codes, $S$ will be the set of all
possible packets, $n=\log |S|$ the length of each packet, and $\ell$ the number
of packets one codeword contains. The source maps information sequence of length
$k$ bits to a codeword which is a set consisting of $\ell$ packets of length $n$
bits each, and sends these $\ell$ packets through a channel. In the channel,
these packets are permuted, some of them are deleted, some of them are received
erroneously, and possibly some new packets are inserted by a malicious user.
The receiver collects all these packets and attempts to reconstruct the codeword
which was sent, i.e., the information sequence which corresponds to this codeword.
\par We next prove a simple, but basic fact about the correcting capabilities of
subset codes.
\begin{theorem}
\label{thm}
  Assume that a code $\mathcal C$ of type $[n,k,d;\ell]$ is used for transmitting
  packets over a network. Then any pattern of $t$ errors, $\rho$ deletions, and $s$
  insertions can be corrected by the minimum distance decoder (with respect to the
  subset metric), as long as $2(\rho+2t+s)<d$.
\end{theorem}
\begin{proof}
  Let $X\in {\mathcal C}$ be the set/codeword which is transmitted through a channel.
  Let $Y$ be the received set. If $\rho$ packets from $X$ have been deleted, and $s$
  new packets have been inserted, then we easily deduce that $|X\cap Y|\geq |X|-\rho$
  and $|Y|\leq |X|-\rho+s$. Observe further that errors can be regarded as combinations
  of deletions and insertions. Namely, an erroneous packet can be thought of as being
  inserted, while the original packet has been deleted. Therefore, the actual number
  of deletions and insertions is $\rho+t$ and $s+t$, respectively. We therefore
	conclude that $|X\cap Y|\geq |X|-\rho-t$ and $|Y|\leq |X|-\rho+s$, and so
  \begin{equation}
     d(X,Y) = |X| + |Y| - 2|X\cap Y| \leq \rho + 2t + s.
  \end{equation}
  Now, if $2(\rho + 2t + s)<d$, then $d(X,Y)\leq\lfloor\frac{d-1}{2}\rfloor$
  and hence $X$ can be recovered from $Y$.
\end{proof}
\par If only deletions can occur in the channel, we will have $d(X,Y)=\rho$ and a
sufficient condition for unique decodability will be $\rho\leq\lfloor\frac{d-1}{2}\rfloor$.
\par As Theorem \ref{thm} establishes, large enough minimum distance $d$ ensures
that the sent codeword can be recovered for a certain level of channel impairments.
Therefore, this parameter is determined by the channel statistics, i.e., probabilities
of packet error/deletion/insertion, and packet delivery requirements (e.g., error
probability). Other code parameters, $\ell$ and $n$, are also determined by certain
delivery requirements, such as delay, and by the properties of the network, such as
the maximal packet length. A general method for the construction of subset codes with
specified parameters, which reduces to the construction of binary codes, is described
in Section \ref{isometry}. Another simple method, via packet-level block codes and
sequence numbers, is illustrated in the following subsection.
\subsection{Examples of subset codes}
\label{examples}
In this subsection, we give a simple example of subset codes to illustrate the
above definitions.
\par How does one encode information in a set? One possible solution (which
is widely used in practice) is to add a sequence number to every packet sent,
thus achieving resilience to arbitrary permutations. To illustrate this, assume
that the source has two packets to send, $p_0$ and $p_1$. Note that, from the
point of view of the receiver, the sequence $(p_0, p_1)$ is not the same as
the sequence $(p_1, p_0)$; these two sequences carry different information.
In the permutation channel, however, either of these two sequences can be
received when $(p_0, p_1)$ is sent. The sender therefore sends $(q_0, q_1)$
instead, where $q_i=(i,p_i)$ is the new packet formed by prepending a sequence
number to the packet $p_i$. Note that sequences $(q_0, q_1)$ and $(q_1, q_0)$
are now identical to the receiver because in both cases it will extract $(p_0, p_1)$
and further process these packets. This means that the carrier of information
is actually a \emph{set} $\{q_0, q_1\}=\{(0,p_0),(1,p_1)\}$. This approach,
combined with some classical packet-level error-correcting code, provides an
example of subset codes that we describe next.
\par Let $\mathcal A$ be the set of all packets the source can possibly send.
Assume that $|{\mathcal A}|=2^m$, so that we can think of information packets
as having $m$ bits. Assume further that the source wishes to send $k$ such
packets, $p_0,\ldots,p_{k-1}$ to a destination over a network, i.e., over a
permutation channel with errors, deletions, and insertions. To protect the
packets the source defines some packet-level block code (see, e.g., \cite{rizzo}),
and uses the corresponding encoder to map these $k$ packets to $\ell>k$ packets,
$q_0,\ldots,q_{\ell-1}$. To cope with the permutations in the channel, the source
further adds a sequence number of length $\log_2 \ell$ bits\footnote{\,For notational
simplicity we disregard the fact that the actual length is $\lceil\log_2 \ell\rceil$.}
to every packet $q_i$. This gives a subset code of type $[m+\log_{2}\ell,km,d;\ell]$,
where $d$ is its minimum distance whose concrete value is irrelevant for this example.
In words, the length of the packets is $m+\log_2\ell$ bits, there are $2^{km}$
possible information sequences (and hence the same number of codewords), and each
codeword consists of $\ell$ packets. The rate of the code is therefore
$R=\frac{km}{\ell(m+\log_{2}\ell)}$.
\par To further clarify the above arguments, assume that the Reed-Solomon (RS) code
is used as a packet-level block code in the above scenario. Namely, the message
to be sent ($k$ packets, $p_0,\ldots,p_{k-1}$, of length $m$ bits each) is being
regarded as a polynomial of degree at most $k-1$ over $\mathbb{F}_{2^m}$:
\begin{equation}
  u(z) = \sum\nolimits_{i=0}^{k-1} p_i z^i.
\end{equation}
The codeword represents the sequence of evaluations of this polynomial in $\ell$
fixed different points in $\mathbb{F}_{2^m}$. Denote these points by
$\alpha_0,\ldots,\alpha_{\ell-1}$, so that the codeword is
$u(\alpha_0),\ldots,u(\alpha_{\ell-1})$. The resulting code
has minimum (Hamming) distance $\ell-k+1$ \cite{sloane}. Now, $u(\alpha_i)$'s are
being treated as packets (these are the $q_i$'s from the previous paragraph), and
each packet is being added a sequence number $i$ (index of the point of evaluation
of the message polynomial). As already explained, these sequence numbers enable
the receiver to recover from permutations, but also from deletions and insertions
because it can keep track of evaluation points. Finally, the codeword corresponding
to the information sequence $(p_0,\ldots,p_{k-1})$ is a set
$U=\left\{(i,u(\alpha_i)):i=0,\ldots,\ell-1\right\}$. Since two polynomials $u$ and
$v$ of degree $k-1$ can agree on at most $k-1$ different points, we conclude that
$|U\cap V|\leq{k-1}$ and therefore $d(U,V)\geq{2(\ell-k+1)}$. Thus, we have defined
a constant-cardinality subset code of type
$\left[m+\log_2 \ell,km,2(\ell-k+1);\ell\right]$, and rate:
\begin{equation}
  R = \frac{km}{\ell\left(m+\log_2 \ell\right)}.
\end{equation}
This code is a subset analog of the K\"otter-Kschischang subspace code \cite{kk}
designed for RLNC networks.
\par Even though RS codes are maximum distance separable \cite{sloane}, the subset
codes obtained in this way are not. Namely, adding a sequence number is not an
optimal way of encoding information in a set (though this suboptimality is not
a concern in practice for sufficiently large packet lengths $m$, because sequence
numbers only take a couple of bytes in the packet header). The other reason for
non-optimality is that these codes are constant-cardinality codes; larger codes
can be obtained if one allows codewords of different cardinality. This is analogous
to the relation of general subspace codes in projective spaces and constant-dimension
codes \cite{etzion}. In the following section we discuss how one can construct optimal
(in any sense) subset codes.
\par As a final note here we point out that the codes constructed in this way
(via packet-level block codes and sequence numbers) are, to the best of our
knowledge, the only type of error-correcting codes for the permutation channel
described in the literature (see, for example, the construction of the ``outer"
code in \cite{schul}). As established above, they are in fact only a special
case of subset codes.
\section{Subset codes as binary codes}
\label{isometry}
Let $S$ be a nonempty finite set with some implied ordering of its elements,
and observe the space $\{0,1\}^{|S|}$ of all binary sequences of length $|S|$
(denoted also $2^S$). Each binary sequence $\boldsymbol{x}\in2^S$ defines a
subset $X\subseteq S$ containing elements defined by the positions of ones in
$\boldsymbol{x}$. As is well-known, this mapping of subsets to binary sequences
is an isomorphism between groups $\left({\mathcal P}(S),\bigtriangleup\right)$
and $\left(2^S,\oplus\right)$, where $\oplus$ denotes the XOR operation (addition
modulo 2). Furthermore, it is easy to show that the Hamming distance between
two sequences $\boldsymbol{x},\boldsymbol{y}\in 2^S$ is precisely the subset
distance between the corresponding subsets $X,Y\subseteq S$:
\begin{equation}
 d_\textsc{h}(\boldsymbol{x},\boldsymbol{y})
                              = w_\textsc{h}(\boldsymbol{x}\oplus\boldsymbol{y})
                              = |X\bigtriangleup Y|
                              = d(X,Y),
\end{equation}
where $w_\textsc{h}$ denotes the Hamming weight of a sequence. In
other words, this mapping is also an isometry between metric spaces
$\left({\mathcal P}(S),d\right)$ and $\left(2^S,d_\textsc{h}\right)$. This
means that the subset codes in fact represent only another way to look at
classical codes in the binary Hamming space, and vice versa. In other words,
\emph{the study of subset codes and their properties reduces to the well-known
theory of binary codes}. Constant-cardinality codes are then equivalent to
constant-weight binary codes. Finally, we note that the classical binary
codes corresponding to $[n,k,d;\ell]$ subset codes have parameters
$\left(2^n,k,d\right)$.
\par The above reasoning, though quite elementary, has an important implication.
It shows that classical codes developed for binary channels (such as the Binary
Symmetric Channel) define in a very natural way codes for correcting errors,
deletions, and insertions in networks. Consequently, many familiar constructions
of binary codes can be applied to subset codes. Namely, once the code parameters
are determined from the given channel statistics and packet delivery requirements,
the subset code with these parameters can be constructed via the corresponding
binary code. For example, a constant-cardinality $[n,k,d;\ell]$ code could be
designed as a constant-weight binary code with codeword weights $\ell$ and
parameters $\left(2^n,k,d\right)$, as explained above.
\par The following toy example illustrates the above notions.
\begin{example}
  Let $S=\{a,b,c,d\}$. Any subset of $S$ can be identified by a binary sequence
  of length $4$; for example $\{a,b\}\leftrightarrow 1100$, $\{b,d\}\leftrightarrow 0101$,
  etc. Consider now some code in $\{0,1\}^4$, e.g., ${\mathcal C}=\{1100, 1010, 0110, 0011\}$.
  The subset counterpart of this code is then ${\mathcal C}_\textsc{s}=\left\{\{a,b\},\{a,c\},\{b,c\},\{c,d\}\right\}$. 
  The distance between two subsets of $S$ is the Hamming distance between the
  corresponding binary sequences, for example:
  \begin{equation}
    d\left(\{a,b\},\{a,c\}\right) = \left|\{b,c\}\right| = 2 = d_\textsc{h}(1100,1010)
  \end{equation}
  so that all properties of $\mathcal C$ directly translate into equivalent
  properties of the subset code ${\mathcal C}_\textsc{s}$. The code
  ${\mathcal C}_\textsc{s}$ is a constant-cardinality code of type $[2,2,2;2]$.
\end{example}
\par The above example can be extended to arbitrary sets $S$ and binary codes
${\mathcal C}\subseteq 2^S$.
\section{Some practical considerations}
\label{comments}
In this section, we give some comments on subset codes and the channel
model that could be relevant for their analysis in practical scenarios.
\subsubsection*{Comments on binary codes}
One constraint on the binary codes corresponding to $[n,k,d;\ell]$ subset
codes should be pointed out. Namely, ``practical" subset codes will certainly
require that $\ell\ll2^n$, i.e., that the number of packets in one codeword
is much smaller than the number of all possible packets. This means that binary
codes corresponding to (practically feasible) subset codes will only have
small weight codewords. Moreover, the fact that binary codes corresponding to
$[n,k,d;\ell]$ subset codes have exponential length ($2^n$) places additional
complexity constraints on the code design.
\subsubsection*{Comments on the channel model}
The links in networks can generally be unreliable. For example, if a large
packet is sent over a wireless link, it is highly probable that it will be
hit by an error, i.e., that at least one of its bits/symbols will be received
incorrectly. Furthermore, this error probability increases with the packet
length $n$. In such a scenario it can happen (with fairly high probability)
that all of the packets from the sent codeword are erroneous, in which case
$X\cap Y=\emptyset$ and reliable recovery is impossible. Subset codes alone
do not provide a good protection from errors in such cases. One way to solve
this problem is to additionally  protect each packet with its own error
correcting code. This solution is in agreement with
current networking practice. Namely, as already noted, we treat here an
end-to-end network model and hence assume that (subset) coding is done on
the transport or application layer. In most networks, packets on lower layers
(e.g., link and physical layer) include some error correcting/error detecting
codes (such as LDPC codes for error correction combined with CRC codes for
error detection). These codes effectively create a channel that we treat here,
namely, they keep the link-layer packet error probability at a ``reasonable"
level.
\par Packet insertions also deserve a comment regarding possible practical
applications of subset codes. In general, by inserting enough packets an
adversary can always prevent the receiver from correctly decoding the received
set. Thus we also assume in our model that the number of insertions is relatively
small, or at least that it behaves as a random variable whose parameters we
can estimate and then design the code with respect to this estimated channel
statistics. This may not be the case in practice because insertions inherently
represent deliberate interference, but our assumption can certainly be achieved
by a proper authentication protocol; that way the receiver will recognize and
disregard (most of) the inserted packets. That is to say that subset codes do
\emph{not} provide any cryptographic protection; insertions are treated here
because they naturally fit in the model, along with deletions and errors.
\par We note that the above comments on errors and insertions are also valid
for subspace codes in network coded networks.
\section{Conclusion}
\label{conclusion}
We have presented a different view on the problem of forward error correction in
the packet permutation channels. One advantage of the presented approach is that
it unifies to some extent coding for RLNC networks and routed packet networks. We
have introduced subset codes as appropriate constructs for these channels. Some
basic properties of subset codes have been established, the most interesting of
which is their equivalence to the classical binary codes.
%
%
%

%
%

\begin{thebibliography}{9}
%
\bibitem{galag}
   D. P. Bertsekas and R. Gallager,
   \emph{Data Networks},
   2nd edition. Prentice Hall, 1992.
\bibitem{kk}
   R. K\"{o}tter and F. R. Kschischang,
   ``Coding for Errors and Erasures in Random Network Coding,"
   \emph{IEEE Trans. Inf. Theory}, vol. 54, no. 8, pp. 3579--3591, Aug. 2008.
\bibitem{schul}
   L. J. Schulman and D. Zuckerman,
   ``Asymptotically Good Codes Correcting Insertions, Deletions, and Transpositions,"
   \emph{IEEE Trans. Inf. Theory}, vol. 45, no. 7, pp. 2552--2557, Nov. 1999.
\bibitem{lattice}
   G. Birkhoff,
   \emph{Lattice Theory},
   3rd edition. American Mathematical Society, 1967.
\bibitem{silva}
   D. Silva and F. R. Kschischang,
   ``On Metrics for Error Correction in Network Coding,"
   \emph{IEEE Trans. Inf. Theory}, vol. 55, no. 12, pp. 5479--5490, Dec. 2009.
\bibitem{rizzo}
   L. Rizzo,
   ``Effective Erasure Codes for Reliable Computer Communication Protocols,"
   \emph{ACM SIGCOMM Computer Communication Review}, vol. 27, no. 2, pp. 24--36, Apr. 1997.
\bibitem{sloane}
   F. J. MacWilliams and N. J. A. Sloane,
   \emph{The Theory of Error-Correcting Codes}, 
   North-Holland Publishing Company, 1977.
\bibitem{etzion}
   T. Etzion and A. Vardy,
   ``Error-Correcting Codes in Projective Space,"
   \emph{IEEE Trans. Inf. Theory}, vol. 57, no. 2, pp. 1165--1173, Feb. 2011.
%
\end{thebibliography}
\end{document}